\documentclass[12pt]{article}

\usepackage{latexsym}
\usepackage{amssymb,amsfonts,amsmath}
\usepackage{graphicx} 
\usepackage{indentfirst}
\usepackage{bbm}
\usepackage{amssymb}
\usepackage{verbatim}
\usepackage{amsmath, amsthm,amssymb}
\usepackage{mathrsfs}
\usepackage{hyperref}
\usepackage{amsfonts}
\usepackage{dsfont}
\usepackage{cite}
\usepackage{xcolor}
\usepackage[multiple]{footmisc}

\parskip=\medskipamount

\newcommand {\cM}{{\cal M}}
\newcommand {\cN}{{\cal N}}

\def\a{\alpha}
\def \bi{\bibitem}

\def\b{\beta}
\def\c{\chi}
\def\d{\delta}

\def\f{\phi}
\def\g{\gamma}
\def\G{\Gamma}

\def\l{\lambda}

\def\o{\omega}

\def\s{\sigma}

\def\F{\Phi}
\def\J{\Psi}

\def\S{\Sigma}

\def\rd{{\rm d}}
\def\ri{{\rm i}}
\def\re{{\rm e}}

\newcommand{\ad}{{\dot{\alpha}}}                           
\newcommand{\ve}{\varepsilon}                            

\newcommand{\pa}{\partial}                           
\newcommand{\hf}{\frac12}

\newcommand{\vf}{\varphi}

\newcommand{\be}{\begin{equation}}
\newcommand{\ee}{\end{equation}}
\newcommand{\bea}{\begin{eqnarray}}
\newcommand{\eea}{\end{eqnarray}}

\newcommand{\bm}[1]{\mbox{\boldmath$#1$}}

\def\double #1{#1{\hbox{\kern-2pt $#1$}}}

\newif\ifdtup

\newcommand{\bsubeq}{\begin{subequations}}
\newcommand{\esubeq}{\end{subequations}}

\numberwithin{equation}{section}

\newcommand{\sU}{\mathsf{U}}

\begin{document}

\begin{titlepage}
\begin{flushright}
March, 2023 \\
Revised version: April, 2023
\end{flushright}
\vspace{5mm}

\begin{center}
{\Large \bf A supersymmetric nonlinear sigma model analogue of the ModMax theory
}
\end{center}

\begin{center}

{\bf Sergei M. Kuzenko and I. N. McArthur} \\
\vspace{5mm}

\footnotesize{
{\it Department of Physics M013, The University of Western Australia\\
35 Stirling Highway, Perth W.A. 6009, Australia}}  
~\\
\vspace{2mm}
~\\
Email: \texttt{ 
sergei.kuzenko@uwa.edu.au, ian.mcarthur@uwa.edu.au}\\
\vspace{2mm}

\end{center}

\begin{abstract}
\baselineskip=14pt
A decade ago, it was shown that associated with any model for $\mathsf{U}(1)$ duality-invariant nonlinear electrodynamics there is a unique $\mathsf{U}(1)$ duality-invariant supersymmetric nonlinear sigma model formulated in terms of chiral and complex linear superfields. Here we study the ${\cal N}=1$ superconformal $\sigma$-model analogue of the conformal duality-invariant electrodynamics known as the ModMax theory. We derive its dual formulation in terms of chiral superfields and show that the target space is a K\"ahler cone with $\mathsf{U}(1)\times \mathsf{U}(1)$ being the connected component of the isometry group.
\end{abstract}
\vspace{5mm}

\vfill

\vfill
\end{titlepage}

\newpage
\renewcommand{\thefootnote}{\arabic{footnote}}
\setcounter{footnote}{0}


\allowdisplaybreaks


\section{Introduction}

Ten years ago, we proposed a family of $\sU(1)$ duality-invariant $\cN=1$ supersymmetric nonlinear sigma models \cite{KMcA}. Such a $\s$-model
is realised in terms of a chiral scalar $\F$, a complex linear  scalar $\S$, and their conjugates,  $\bar \F $ and $\bar \S$,
 \bea
\bar{D}_{\ad} \Phi= 0~, \qquad  \bar{D}^2  \S = 0~.
\label{constraints}
\eea
The action is of the form
\bea
S_{\text{CCL}}= \int \rd^4x \rd^4 \theta \, L(\Phi, \bar{\Phi}, \S, \bar{\S})~.
\label{1.1}
\eea
The equations of motion for $\S$ and $\F$ are 
\bea
 \bar{D}_{\ad} \, \frac{ \partial L}{\partial \S} = 0~, \qquad 
 \bar{D}^2 \, \frac{\partial L}{\partial \Phi} = 0~. 
\eea
We see that the equations of motion have the same functional form as the off-shell 
constraints but with $\F$  and $\S$ interchanged.
As a result, ``duality'' rotations that mix $\Phi$ and 
${ \partial L}/{\partial \S}$,
 and $\S$ 
and ${ \partial L}/{\partial \Phi},$ 
leave the constraints and the equations of motion invariant.

With the notation $ X := ( \F, \bar{\F}, \S , \bar{ \S}) $, the 
$\sU(1)$ duality-invariant supersymmetric nonlinear $\s$-models are characterised 
by continuous duality rotations of the form 
\begin{subequations}\label{2.3}
\bea
\left(
\begin{array}{c}
\Phi' \\  \frac{\partial L(X')}{\partial \S'}
\end{array}
\right)
&=& \left(
\begin{array}{cc}
\cos \l & \, \, \sin \l   \\  - \sin \l & \, \, \cos \l 
\end{array}
\right) \, \, \left(
\begin{array}{c}
\Phi\\  \frac{\partial L(X)}{\partial \S}
\end{array}
\right)  
\eea
and
\bea
\left(
\begin{array}{c}
\S' \\  \frac{\partial L(X')}{\partial \Phi'}
\end{array}
\right)
&=& \left(
\begin{array}{cc}
\cos \l & \, \, \sin \l   \\  - \sin \l & \, \, \cos \l
\end{array}
\right) \, \, \left(
\begin{array}{c}
\S \\  \frac{\partial L(X)}{\partial \Phi}
\end{array}
\right)~.
\eea
\end{subequations}
As demonstrated in \cite{KMcA}, duality invariance of the theory implies that 
the Lagrangian $ L(X) =L(\Phi,\bar{\Phi},  \S, \bar{\S}) $ 
must obey the differential equation
\bea
0 = \Phi \, \S +    \bar{\Phi} \, \bar{\S} 
 +  \frac{\partial L }{\partial \Phi} \, \frac{\partial L }{\partial \S}
+  \frac{\partial L  }{\partial \bar{\Phi}}  \, \frac{\partial L }{\partial \bar{\S}} 
 ~.
\label{SS}
\eea

Any nonlinear $\s$-model of the form (\ref{1.1}), which involves 
chiral and complex linear (CCL) superfields,  
has a purely chiral formulation which is obtained by performing a 
superfield Legendre transformation that dualises the complex linear 
superfield $\S$ and its conjugate $\bar \S$ into a chiral scalar $\J$ 
and its conjugate $\bar \J$. 
It is worth recalling its derivation. 
Starting from the $\s$-model \eqref{1.1}, we introduce a first-order action
\be
S_{\text{first-order}}= \int \rd^4x \rd^4 \theta \, \Big\{ L(\Phi, \bar{\Phi}, \S, \bar{\S})
 + \Psi \S + \bar{\Psi} \bar{\S} \Big\}~,
\label{1.6}
\ee
where $\S$ is taken to be an unconstrained complex superfield,
while the Lagrange multiplier $\J$ is chosen to be chiral, 
\bea
\bar D_\ad \J =0~.
\eea
The original CCL $\s$-model \eqref{1.1} is obtained from \eqref{1.6} by integrating out 
the Lagrange multipliers $\J$ and $\bar \J$. Instead, we can integrate out 
the auxiliary superfield $\S$ and its conjugate $\bar \S$ using the  corresponding 
equation of motion 
\be
 \frac{\partial L}{\partial \S } + \Psi =0 
\label{eom}
\ee
and the  conjugate equation. This leads to the chiral formulation
\bea
S_{\text{chiral}}= \int \rd^4x \rd^4 \theta \, K(\Phi, \bar{\Phi}, \Psi, \bar{\Psi}) ~,
\label{sigma-model_chiral}
\eea
where the corresponding Lagrangian $K$ is the Legendre transform of $L$,
\be
K(\Phi, \bar{\Phi}, \Psi, \bar{\Psi}) 
= L(\Phi, \bar{\Phi}, \S, \bar{\S}) + \Psi \S + \bar{\Psi} \bar{\S}~.
\label{LD}
\ee
The dual Lagrangian $K(\Phi, \bar{\Phi}, \Psi, \bar{\Psi})$ 
is  the K\"ahler potential for a K\"ahler target space. 

In the chiral formulation, the condition of $\sU(1)$ duality invariance \eqref{SS} was shown  \cite{KMcA} to become 
the requirement that $K(\Phi, \bar{\Phi}, \Psi, \bar{\Psi})$ is invariant  under rigid $\sU(1)$ transformations
\bea
\d \Phi = - \l \Psi~, \quad \d \Psi =  \l \Phi~, \qquad \l \in {\mathbb R}~.
\label{1.11}
\eea
Thus  the isometry group of the target K\"ahler space 
is nontrivial, since it contains the $\sU(1)$ subgroup of  transformations \eqref{1.11}.

An important class of duality-invariant nonlinear $\sigma$-models \eqref{1.1} is described
by a Lagrangian of the form 
\bea
L(\Phi, \bar{\Phi}, \S, \bar{\S}) = L(\o, \bar \o)~,
\eea
where the complex variable $\o$ is defined by
\be 
\omega = \bar{\Phi} \Phi  - \bar{\S} \S  + \ri \, (\Phi  \S + \bar{\Phi} \bar{\S}) = (\Phi  + \ri \bar{\S}) \, (\bar{\Phi}  + \ri \S) ~.
\label{omega}
\ee
For such $\s$-models, the condition for duality invariance, eq. \eqref{SS}, was shown 
\cite{KMcA} to take  the form
\bea 
{\rm Im} \, \left\{ \omega 
- 4 \, \omega \left( \frac{\partial L}{\partial \omega} \right)^2  \right\} =0
~ .
\label{universal}
\eea

Building on the influential 1981 work by Gaillard and Zumino \cite{GZ1},
the general theory of $\sU(1)$ duality-invariant
models for nonlinear electrodynamics in four dimensions was developed in the mid 1990s \cite{GR1,GR2,GZ2,GZ3} and the early 2000s \cite{IZ_N3,IZ1,IZ2} (see also \cite{IZ3}), 
including the case of duality-invariant theories with higher derivatives \cite{KT2}.
In general, 
nonlinear electrodynamics is described by a Lagrangian of the form
\bea
L_{\rm NLED} (F_{ab}) = L({\bm \o}, \bar {\bm \o})~, \label{1.15}
\eea
where we have introduced 
\bea
-{\bm \o} := F_{\a \b} F^{\a \b} = \frac 14 F^{ab} F_{ab}+ \frac{\ri}{4} \, F^{ab} \tilde{F}_{ab} ~,
\label{bmo}
\eea
with $F_{ab}$ being the electromagnetic field strength.
The Gaillard-Zumino-Gibbons-Rasheed  
condition for $\sU(1)$ duality invariance of nonlinear electrodynamics \eqref{1.15}
is 
\bea
\widetilde{G}^{ab}G_{ab}  +  \widetilde{F}^{ab}F_{ab} = 0~,
\qquad
\widetilde{G}^{ab} (F):=
\hf \, \ve^{abcd}\, G_{cd}(F) =
2 \, \frac{\pa L(F)}{\pa F_{ab}}~, 
\label{SDE}
\eea
originally given in \cite{GR1,GZ2,GZ3}.\footnote{Actually, the self-duality equation  was derived for the first time by Bialynicki-Birula \cite{B-B}, but unfortunately his work was largely unnoticed.
}
As shown in \cite{KT2}, this condition for $\sU(1)$ duality invariance of nonlinear electrodynamics can be expressed in the form of  \eqref{universal} with a $\o$ replaced by $ {\bm \o}. $ 
This means that any $\sU(1)$ duality-invariant Lagrangian for nonlinear electrodynamics will also generate a $\sU(1)$ duality-invariant supersymmetric $\s$-model by replacing $ {\bm \o} $ defined in \eqref{bmo} by $\o$ defined in \eqref{omega}.

Three years ago, a unique conformal $\sU(1)$ duality-invariant nonlinear electrodynamics was constructed \cite{BLST} (see also \cite{Kosyakov})
and called the ModMax theory. It is described by the Lagrangian
 \bea
L_{\rm conf}({\bm \o}, \bar{\bm \o}) = \frac12  ({\bm \o} + \bar{\bm \o}) \cosh \g  
+ \, \,\sqrt{ {\bm \o} \,  \bar{\bm \o}  } \sinh \g ~ ,
\label{1.17}
\eea
with $\g $ a coupling constant.
This model does not possess a weak-field expansion, which is why such theories had not been considered earlier.\footnote{Using the Ivanov-Zupnik formalism
 \cite{IZ_N3,IZ1,IZ2,IZ3}, this theory was re-derived in Appendix A of \cite{K2021}.} The $\cN=1$ supersymmetric extension of the 
ModMax theory \eqref{1.17}
was constructed in \cite{BLST2}, and alternative derivations of the resulting theory were given in \cite{K2021} using the approaches advocated in \cite{KT2,KT1,K13}.
It should be mentioned that every
$\mathsf{U}(1)$ duality-invariant 
nonlinear electrodynamics described by the relations
\eqref{1.15} and \eqref{SDE}
is contained in a  $\mathsf{U}(1)$ duality-invariant model for the $\cN=1$ vector multiplet proposed in \cite{KT2,KT1}. 


\section{Chiral formulation}

Our aim in this paper is to study the supersymmetric $\s$-model analogue\footnote{Inspired by the Australian cinematographic tradition, 
it seems suitable to call \eqref{2.1} the MadMax $\s$-model, although we will not pursue this terminology.} 
 of the ModMax theory \eqref{1.17},
\bea
L(\o, \bar{\o}) = \frac12 (\o + \bar{\o}) \cosh \g   + \sqrt{ \o \,  \bar{\o}  }\sinh \g  ~,
\label{2.1}
\eea
where $\o$ is defined by \eqref{omega}. 
In terms of the original dynamical variables $ X = ( \F, \bar{\F}, \S , \bar{ \S}) $, the Lagrangian 
is given by 
\bea
L(X) =    (\F \bar{\F} - \S \bar{\S}) \cosh \g 
+  
\sqrt{ ( \F^2 + \bar{\S}^2) (\bar{\F}^2 + \S^2 )} \sinh \g  ~.
\label{2.2}
\eea
It is a superconformal field theory. In the framework of \cite{Kuzenko:2007qy},
the superconformal transformation laws of $\F$ and $\S$ are given by eq. (6.1).

 As discussed in the Introduction, we can perform a Legendre transformation to dualise the complex linear superfield $\S$ and its conjugate $ \bar{\S}$ into a chiral scalar $\J$ and its conjugate $\bar{\J}$.  In terms of the superfields $ X_D := ( \F, \bar{\F}, \J ,  \bar{ \J}), $ the dual formulation is determined by the action
\bea
L_D(X_D) = L(X) + \J \S + \bar{\J} \bar{\S} . ~
\eea
The equation of motion $\frac{ \partial L_D(X_D)}{ \partial \S} = 0 $ yields
\bea
\J & = & \bar{\S} \,  \cosh \g  - \frac{\ri}{2} ( \F + \ri \bar{\S})  \left( \frac{\bar{\o}}{\o} \right)^{\frac12} \, \sinh \g  + \frac{\ri}{2} ( \F - \ri \bar{\S})  \left( \frac{\o}{\bar{\o}} \right)^{\frac12} \,  \sinh \g  \\
&= &  \bar{\S} \, \cosh \g  -  \S \left(\frac{ \F^2 + \bar{\S}^2}{\bar{\F}^2 + \S^2 } \right)^{\frac12} \, \sinh \g  . ~
\eea
Defining 
\bea
X =  \left(\frac{ \F^2 + \bar{\S}^2}{\bar{\F}^2 + \S^2 } \right)^{\frac12}
\label{X}
\eea
and noting that $\bar{X} = \frac{1}{X},$
\bea
\begin{pmatrix}
\J \\
\bar{\J}
\end{pmatrix} 
= 
\begin{pmatrix}
\cosh \g & - X \sinh \g \\
- \frac{1}{X} \sinh \g &  \cosh \g
\end{pmatrix}
\,
\begin{pmatrix}
\bar{\S} \\
\S
\end{pmatrix} .
\eea
Inverting this relation,
\bea
\begin{pmatrix}
\bar{\S} \\
\S
\end{pmatrix} 
= 
\begin{pmatrix}
\cosh \g &  X \sinh \g \\
 \frac{1}{X} \sinh \g &  \cosh \g
\end{pmatrix}
\,
\begin{pmatrix}
\J \\
\bar{\J}
\end{pmatrix} . \label{Sigma}
\eea

Substituting the expression above for $ \S $ and $ \bar{ \S } $ into the definition (\ref{X}) of $X$, one obtains the quadratic equation 
\bea
X^2 = \frac{\F^2 + \J^2}{ \bar{\F}^2 + \bar{ \J}^2 } \quad   \Rightarrow  \quad X = \pm \left(  \frac{\F^2 + \J^2}{ \bar{\F}^2 + \bar{ \J}^2 } \right)^{\frac12}. ~
\eea
This therefore allows us, using equation ( \ref{Sigma}), to express the complex linear superfields $\S$ and $\bar{\S}$ in terms of the chiral superfields $\F$, $ \bar{\F}$, $\J$ and $ \bar{\J},$
as required.

In order to compute the dual Lagrangian $L_D (X_D)$, we note that using the original definition (\ref{X}) of $X$, we can write
\bea
\sqrt{ \o \, \bar{\o} } = X \, (\bar{\F}^2 + \S^2),
\eea
and equivalently,
\bea 
\sqrt{ \o \, \bar{\o} } = \frac{1}{X} \, (\F^2 + \bar{\S}^2).
\eea
Thus we can express $ \sqrt{ \o \, \bar{\o} } $ in the symmetric form 
\bea
\sqrt{ \o \, \bar{\o} } = \frac{X}{2} \, (\bar{\F}^2 + \S^2) + \frac{1}{2 X} \, (\F^2 + \bar{\S}^2).
\eea
Substituting this into
\bea
L_D (X_D) =   \frac12  (\o + \bar{\o}) \, \cosh \g  + \sqrt{ \o \,  \bar{\o}  } \,  \sinh \g  + \J \S + \bar{\J} \bar{\S} ~
\eea
and using the expressions (\ref{Sigma}) for $ \S$ and $\bar{ \S}$, 
we can obtain the K\"ahler potential
 $ L_D (X_D)= K \big(\F, \bar \F, \J , \bar \J \big) $.
The result is 
\bea
K \big(\F, \bar \F, \J , \bar \J \big)=    ( \F \, \bar{\F}  + \J \, \bar{ \J} ) \cosh \g+    
\sqrt{ (\F^2 + \J^2) \, ( \bar{ \F}^2 + \bar{ \J}^2 )} \sinh \g~.
\label{2.14}
\eea

The target space of the $\s$-model is a K\"ahler cone, following the terminology of \cite{GR}, in particular the K\"ahler potential obeys the homogeneity condition
\bea
\Big( \F \frac{\pa}{\pa \F}  +  { \J}   \frac{\pa }{\pa { \J} }
 \Big) K \big(\F, \bar \F, \J , \bar \J \big)
= K \big(\F, \bar \F, \J , \bar \J \big)~.
\label{homo5}
\eea
With the notation $\f^i = (\F, \J)$ and ${\bar \f}^{\bar i} = (\bar \F , \bar \J)$, 
the K\"ahler metric $g_{i\bar j} (\f, \bar \f)$ is 
\bea \label{metric}
g=( g_{i \bar j}) &=& {\mathbbm 1}_2 \cosh \g 
+ \frac{\f \, \f^{\dagger} }{ \sqrt{ \f^{\rm T} \f \f^\dagger \bar \f }} \sinh \g~,
\eea
where $\f$ is viewed as a column-vector. 
A short  calculation gives
\bea
\det (g_{i \bar j}) = \cosh^2\g + \frac{ \F \bar \F + \J \bar \J} 
{ \sqrt{ (\F^2 + \J^2) \, ( \bar{ \F}^2 + \bar{ \J}^2 )} } \cosh \g \sinh \g ~.
\label{metric-determinant}
\eea
For $\g \neq 0$, the matrix elements of \eqref{metric} are nonsingular  
in the domain $\F^2 + \J^2 \neq 0$, which we identify with the target space $\cM^4$ 
of the $\s$-model. It follows from \eqref{metric} and \eqref{metric-determinant} that 
the metric is positive definite on $\cM^4$ provided  $\g> 0$.  It is interesting that the condition $\g\geq 0$ also naturally occurs for the ModMax theory \cite{BLST}, since for $\g<0$ superluminal propagation becomes possible. 

The connected component of the isometry group $G$ of the K\"ahler cone $\cM^4$
is $\sU(1) \times \sU(1)$. 
It consists of holomorphic transformations  of the form
\bea
\f \to g \f~, \qquad g = \re^{\ri \a} \re^{-\ri \l \s_2}~, \quad \a,\l \in {\mathbb R} ~,
\eea
where $\s_2$ is one of the Pauli matrices $\s_I = (\s_1, \s_2, \s_3)$.
The $\l$-transformation is a finite version of \eqref{1.11}.
The $\a$-transformation is generated by 
the homothetic conformal Killing vector 
\bea 
\c = \c^i \pa_i+ {\bar \c}^{\bar i} \pa_{\bar i}
~, 
\qquad \c^i = \f^i~, \qquad \pa_i = \frac{\pa}{\pa \f^i}~, \quad \pa_{\bar i} = \frac{\pa}{\pa {\bar \f}^{\bar i} }~.
\eea
In general,  a K\"ahler cone possesses a homothetic conformal Killing vector with the properties \cite{GR}
\bea
\nabla_j \c^i =\d_j{}^i ~, \qquad \nabla_{\bar j} \c^i =0~,
\eea
with $\nabla$  the torsion-free covariant derivative on $\cM^4$.
In particular, $\c$ is holomorphic. The properties of $\c $ include the following:
\bea
\c_i := g_{i \bar j} \bar \c^{\bar j} = \pa_i K, \quad \c^i \pa_i K = K 
\quad \implies \quad K = g_{i\bar j} \c^i \bar \c^{\bar j} ~.
\eea
The K\"ahler potential \eqref{2.14} is positive in the domain $\F^2 + \J^2 \neq 0$.

The isometry group $G$ of $\cM^4$ also includes a discrete holomorphic transformation that may be chosen as $g= \s_1$.\footnote{The isometry group $G$ also includes the anti-holomorphic discrete transformation $\f \to \bar \f$.} 
By multiplying $\s_1$ with certain elements of the subgroup 
$\sU(1) \times \sU(1) \subset G$, one observes that $G$ also includes the following group elements: $\pm \s_I$ and $\pm \ri \s_I$.  In particular, 
the isometry group includes the non-abelian quaternion group 
$Q_8 = \{ \pm \ri \s_I, \pm {\mathbbm 1}_2 \}$.

For completeness, we reproduce the component version of  the $\s$-model action \eqref{sigma-model_chiral}, see e.g. \cite{WB} for the technical details:\footnote{We recall that the Christoffel symbols $\G^i_{jk} $ and the curvature tensor
${R}_{i \bar j k  \bar l} $ are given by the expressions
$\G^i_{jk} = g^{i \bar l} \pa_j \pa_k \pa_{\bar l} {K}$ and
$ {R}_{i \bar j k  \bar l} = \pa_i \pa_k \pa_{\bar j} \pa_{\bar l}  {K}
-{g}^{m \bar n} \pa_i \pa_k \pa_{\bar n} {K}  \pa_{\bar j} \pa_{\bar l}  \pa_m {K}$.} 
\begin{align}
S &= -\int \rd^4x\,  \Big\{
      \pa^a \varphi^i g_{i \bar j}\pa_a \bar\varphi^{\bar j}
     + \ri \l^{\alpha i} g_{i \bar j} \nabla_{\alpha \ad} \bar \l^{\ad \bar j}
     - \hat F^i g_{i \bar j} \bar {\hat F}^{\bar j}
     - \frac{1}{4} (\l^i \l^j) (\bar\l^{\bar k} \bar\l^{\bar l}) R_{i \bar k j \bar l}
     \Big\}~.
    \label{2.22}
\end{align}
Here we have defined the component fields of $\f^i$  in the conventional way
\begin{align}
\varphi^i := \phi^i \vert~, \qquad
\l_\alpha^i := \frac{1}{\sqrt 2} D_\alpha \phi^i\vert~, \qquad
F^i := -\frac{1}{4} D^2 \phi^i\vert
\end{align}
and have made use of the complex field
\begin{align}
\hat F^i := F^i - \frac{1}{2} \Gamma^i{}_{jk} \l^j \l^k
\end{align}
which transforms covariantly under holomorphic reparametrisations.
The K\"ahler metric in \eqref{2.22} depends on the physical scalar fields, 
$g_{i\bar j} (\vf, \bar \vf)$.

The formalism of \cite{KMcA} admits a natural extension to $\cN=1$ supergravity. In particular, the superconformal sigma model \eqref{2.2} and its dual \eqref{2.14} can be coupled to conformal supergravity.
\\

\noindent
{\bf Acknowledgements:}\\
We are grateful to Dmitri Sorokin for useful comments and suggestions. 
The work of SMK is supported in part by the Australian 
Research Council, projects DP200101944 and DP230101629.


\begin{footnotesize}

\end{footnotesize}


\end{document}